\documentclass[12pt]{article}
\usepackage{epsfig}

\newcommand{\eq}{\begin{equation}} 
\newcommand{\eqx}{\end{equation}}
\newcommand{\eqn}{\begin{eqnarray}} 
\newcommand{\eqnx}{\end{eqnarray}}

\newcommand{\f}[2]{\frac{#1}{#2}}
\newcommand{\lra}{\longrightarrow}
\newcommand{\ra}{\rightarrow}
\newcommand{\cor}[1]{\left\langle{#1}\right\rangle}

\newcommand{\der}{\partial}
\newcommand{\xpr}{x_\perp}

\newcommand{\zmx}{z_{max}}
\newcommand{\alp}{\alpha'}
\newcommand{\al}{\alpha}
\newcommand{\bt}{\beta}
\renewcommand{\th}{\theta}
\newcommand{\Dl}{\Delta}
\newcommand{\hyp}[4]{{ }_2F_1(#1,#2;#3|#4)}

\newcommand{\sng}{{S_{NG}}}

\newcommand{\var}[1]{\f{\dl \sng}{\dl {#1}}}
\newcommand{\dx}{\partial_x}
\newcommand{\gtt}{h_{\tau\tau}}
\newcommand{\gss}{h_{\sg\sg}}
\newcommand{\dl}{\delta}
\newcommand{\sg}{\sigma}
\newcommand{\atl}{\tilde{A}}

\newcommand{\uzw}[2]{\partial_{#1}\partial_{#2}u}
\newcommand{\du}[1]{\partial_{#1}u}

\newcommand{\zx}{z_x}
\newcommand{\wx}{w_x}

\newcommand{\dlgrav}{\dl_{grav}}
\newcommand{\dlb}{\dl_B}
\newcommand{\dlkk}{\dl_{KK}}
\newcommand{\fal}{\left(\f{a}{L}\right)}

\newcommand{\rr}[4]{#1, {\it #2 \/}{\bf #3} #4}

\begin{document}

\title{High energy scattering and the AdS/CFT correspondence}

\author{R.A. Janik$^{a,b}$  and R. Peschanski$^a$ \\ \\
$^a$Service de Physique Theorique  CEA-Saclay \\ F-91191
Gif-sur-Yvette Cedex, France\\
$^b$M.Smoluchowski Institute of Physics, Jagellonian University\\ Reymonta
4, 30-059 Cracow, Poland}

\maketitle

\abstract{We consider small angle and large impact parameter high
energy scattering of colourless states in SYM
using the AdS/CFT correspondence. The gauge theory scattering
amplitude is linked with a correlation function of tilted Wilson
loops, which can be calculated by the exchange of bulk supergravity
fields between the two corresponding string worldsheets. We identify
the dominant contributions, which all correspond to real phase shifts.
In particular, we find a contribution of the bulk
graviton which gives an unexpected `gravity-like' $s^1$ behaviour of
the gauge theory phase shift in a specific range of energies and
(large) impact parameters.}  

\section{Introduction}

The remarkable duality between supergravity (string theory) on AdS and
supersymmetric gauge theory \cite{adscft} has attracted much attention.
Diverse phenomenae on both sides of the correspondence were investigated
(see e.g. references in \cite{review}).

In this paper we would like to consider the problem of high energy
scattering of massive colourless states, in the large $s$, 
large impact parameter regime (near forward scattering) 
in YM theory, and to study it using the AdS/CFT
correspondence. Although no quantitative predictions are available for
high energy scattering in strongly coupled SYM, there are some
qualitative expectations (unitarity, constraints on the $s$ behaviour
of amplitudes related to analyticity and crossing properties) that
should be satisfied, and which would be 
interesting to investigate from the supergravity side. Apart from
that, in this regime one observes a different hiearchy of
importance of various supergravity fields in comparision to
e.g. `static' quantities like $q\bar{q}$-$q\bar{q}$ potential
\cite{MaldCor}. Another point is that here one expects the
dominant contribution to come just from the `gluonic' sector of the
theory. The contribution of fermions and scalars is expected to be
subleading, and thus the results may not be too closely tied to the
${\cal N}=4$ supersymmetry (see \cite{Bern}).

High energy scattering amplitudes may be parameterized by phase shifts
$\dl$:
\eq
\label{e.dldef}
\f{1}{s} A(s,t)=\int d^2b \;e^{iqb}\;\f{e^{i2\dl(b,s)}-1}{2i}
\quad\quad\quad\quad t=-q^2 
\eqx
where $b$ is the impact parameter (in the following we will denote its
modulus by $L$).
In field theory, there exists a spin-energy relation, the exchange of
elementary scalars gives rise to a phase shift
behaving like $1/s$, vectors lead to $s^0$ behaviour. The exchange of
spin-2 particles (like gravitons) would on the other hand lead to a
dramatic rise $s^1$. This last possibility is excluded in perturbative
SYM with combinations of elementary vector, scalar and fermion
exchanges\footnote{Note, however, that a perturbative resummation at
high energy can lead to the exchange of a compound state and a rise in
the $s$ dependence \cite{BFKL}.}.

Qualitatively one expects the same pattern of behaviour in the AdS/CFT
correspondence. The scattering amplitude will be seen to correspond to
a correlation function of two Wilson loops, which can be evaluated, on
the supergravity side, as the exchange of (bulk) supergravity fields
between the associated string worldsheets. We will show that the same
pattern of spin-energy behaviour persists here and, in particular,
that the graviton 
gives rise to the unexpected $s^1$ dependence in the real part of the
amplitude. Finally we will show that
the Drukker-Gross-Ooguri Legendre transform prescription \cite{neww}
for Wilson loops does not modify this result.

The plan of the paper is as follows. First we discuss general
properties of scattering amplitudes and introduce the appropriate
gauge theory observable. Then we make a passage to euclidean space and
apply the AdS/CFT correspondence. In section 5 we present
calculations for various fields. In section 6 we show that the
Legendre transform prescription for Wilson loops \cite{neww} does not
change the main results. 
Finally we analyze the region of
validity of our calculations and discuss the results.

\section{Scattering amplitudes}

Quark-quark scattering amplitudes in the high energy limit (and small
momentum transfer) can be
conveniently expressed, in the eikonal approximation, in terms of a
correlator of Wilson {\em lines} 
\cite{Nacht,VV,Kor}.
The high energy limit is reached when the lines
move towards the light cone. 
Gauge invariance is restored (see e.g. \cite{VV}) by requiring that the gauge
transformations at both ends of the line are the same. We will here pursue a
different route \cite{Nachtr} by substituting a Wilson {\em loop} for
each of the Wilson lines.
\eq
\label{e.ampinit}
-2is \int d^2\xpr e^{iq\xpr} \cor{\f{W_1W_2}{\cor{W_1}\cor{W_2}}-1}
\eqx
where the Wilson loops follow classical straight lines for quark(antiquark)
trajectories: $W_1\lra x_1^\mu=p_1^\mu\tau\,(+a^\mu)$ and $W_2\lra
x_2^\mu=\xpr^\mu+p_2^\mu\tau\, (+a^\mu)$ and close at infinite times.
This corresponds to the scattering of colorless 
quark-antiquark pairs with transverse separation $a$ (see figure 1 for
the geometry rotated to euclidean space). 

The r{\^o}le of the
quarks in the AdS/CFT correspondence will be played, as in
\cite{Wilson}, by the massive $W$ bosons arising from breaking
$U(N+1)\rightarrow U(N)\times U(1)$.
The geometrical parameters of the configuration can be related to the
energy scales by the relation
\eq
\cosh \chi\equiv \f{1}{\sqrt{1-v^2}}=\f{s}{2m^2}-1
\eqx
where $\chi=\f{1}{2}\log \f{1+v}{1-v}$ is the Minkowski angle
(rapidity) between the two lines, and $v$ is the relative velocity.
Our aim is to apply the AdS/CFT correspondence to calculate
(\ref{e.ampinit}). 
In order to avoid the complications of Lorentzian AdS/CFT
correspondence \cite{lorads} we will link, following \cite{Megg}, this
observable with a related observable in euclidean space. 

\section{Analytical continuation to Euclidean time}

In \cite{Megg} a scattering amplitude with Wilson lines was linked
with an analogous correlator of Wilson lines in Euclidean spacetime
which form an angle $\theta$ (see figure 1). The parametrization of
the Wilson lines is given by euclidean momenta
\eq
\label{e.pareuc}
p_1^E=(1,0,0) \quad\quad\quad\quad 
p_2^E=(\cos\th,-\sin\th,\xpr)
\eqx
After performing the calculation one analytically
continues
\eq
\label{e.analcont}
\th \lra -i\chi\sim -i \log\left(\f{s}{2m^2}\right)
\eqx
to obtain (\ref{e.ampinit}). 

This claim was supported \cite{Megg} by a
number of explicit calculations and a general argument based on
the relation  between correlators of the gauge
fields contracted with momenta both in Euclidean and Minkowski
space\footnote{The same analytical continuation is expected to apply
also to the Wilson loop correlators in the geometry we are considering.}.
A similar link between scattering amplitudes of branes and Euclidean
potentials between branes at angles \cite{braneang} has already been
exploited (see ref. \cite{book}).

\begin{figure}
\centerline{\epsfysize=5cm \epsfxsize=4cm \epsfbox{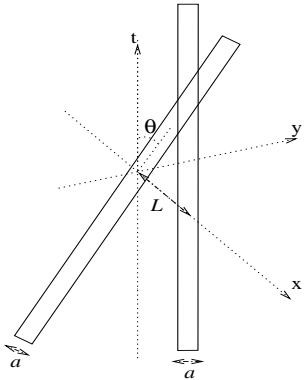}}
\caption{Geometry of the Wilson loops in euclidean space.}
\end{figure}

As an example of this formalism we will consider, following
\cite{MegEik}, the leading order perturbative QED amplitude.
The Wilson line correlator is (for small $e^2$)
\eq
\f{\cor{WW}}{\cor{W}\cor{W}}\propto \cor{e^{ie\int A^\mu
p_{1\mu}^Ed\tau_1} e^{ie\int A^\nu p_{2\nu}^Ed\tau_2}}\sim
e^{e^2\int d\tau_1 d\tau_2 p_{1\mu}^E G_{\mu\nu} p_{2\nu}^E}
\eqx
where $G_{\mu\nu}$ is the (euclidean) photon Green's function (in the
Feynman gauge for simplicity)
\eq
G_{\mu\nu}=\f{1}{4\pi^2} \cdot \f{1}{L^2+(\tau_1-\tau_2 \cos\th)^2
+\tau_2^2 \sin^2\th} g_{\mu\nu}
\eqx
We pass to polar coordinates, perform the integral, drop the infinite
Coulomb phase and obtain
\eq
\exp\left( \f{e^2}{2\pi} \cot \th \cdot \log L \right)
\eqx
Performing the analytical continuation (\ref{e.analcont}) we obtain
immediately the standard QED eikonal result (see e.g. \cite{qedeik},
\cite{Kor}) 
\eq
\label{e.qedeik}
\dl=\f{e^2}{2\pi} \coth \chi \cdot \log L
\eqx

\noindent{}In our case we therefore have to calculate
\eq
\label{e.wcors}
\f{\cor{W_1W_2}}{\cor{W_1}\cor{W_2}}
\eqx
with the Wilson loops parametrized by (\ref{e.pareuc}). Since we want
to perform the calculation nonperturbatively we will apply the
euclidean AdS/CFT correspondence here, and compute the quantity
(\ref{e.wcors}) in the AdS supergravity approximation.

\section{Wilson loop at an angle $\theta$}

The string worldsheet in AdS space, corresponding to a Wilson loop
${\cal C}$ is obtained \cite{Wilson} by minimizing the Nambu-Goto action
\eq
\int_{\partial\Sigma={\cal C}} d\tau d\sg \sqrt{h}
\eqx
where $h_{ab}$ is the induced metric $h_{ab}=G_{\mu\nu} \partial_a
X^\mu \partial_b X^\nu$, $h=\det h_{ab}$ is its determinant  and
$G_{\mu\nu}=(1/z^2)\dl_{\mu\nu}$ the AdS 
metric. For the Wilson loop at an angle $\th$ (see figure 1) we choose the
parametrization\footnote{The $4^{th}$ coordinate on the boundary will
be taken to be a constant.} $\sg\equiv x$,
$t=\tau \cos\th$, $y=\tau \sin\th$, $z=z(x)$, where $z$ is the
$5^{th}$ coordinate of $AdS_5$. This leads to minimizing
the action
\eq
\int dx \f{1}{z^2}\sqrt{1+z_x^2}
\eqx
with no $\th$ dependence. The result \cite{Wilson} is given by 
\eq
z_x=\f{1}{z^2}\sqrt{\zmx^4-z^4}
\eqx
with $\zmx=a\cdot\Gamma(1/4)^2/(2\pi)^{3/2}$, where $a$ is the
transverse size of the loops. The area element on the
worldsheet is given by
\eq
\label{e.area}
d{\cal A}\equiv dt dx \sqrt{h} = dt dx \f{\zmx^2}{z^4} = 
dt dz \f{\zmx^2}{z^2 \sqrt{\zmx^4-z^4}} 
\eqx
For later reference we quote $\gss=\zmx^4/z^6$, $\gtt=1/z^2$ and
$\sqrt{h}=\sqrt{\det h_{ab}}=\zmx^2/z^4$. 

\section{Wilson loop correlators}

\label{s.ww}

The Wilson loop correlators can be computed following
\cite{MaldCor}. One has to distinguish two cases \cite{Gross}. When
the transverse seperation $L$ between the loops is comparable with the
transverse 
size $a$, there may exist a connected minimal surface with the sum
of the two loops as its disjoint boundary (see
e.g. \cite{Zarembo}). However when $L\gg a$ the 
minimal surface has two independent components and in order to
calculate the correlator one has to consider the supergravity
interaction between them as in \cite{MaldCor}. This is the case we
will consider here.
The AdS radius is fixed by convention to 1. Then $\al'=1/\sqrt{4\pi
g_s N}$ and $g^2_{YM}= 2\pi g_s$.

\begin{figure}
\centerline{\epsfysize=5cm \epsfbox{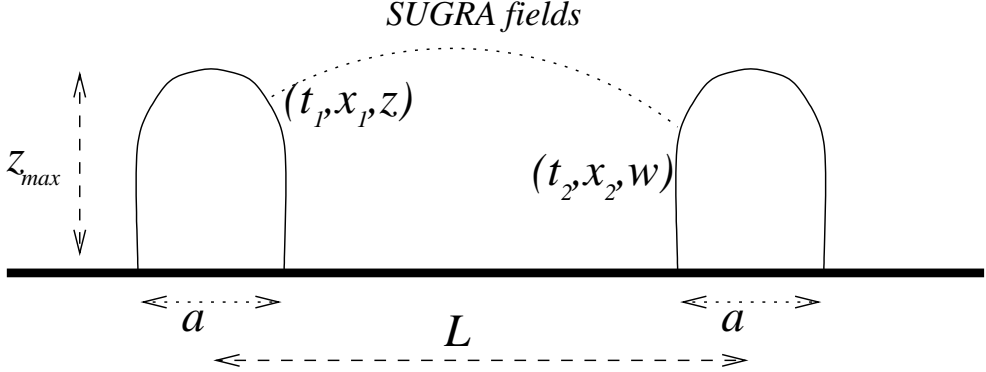}}
\caption{Correlation function of Wilson loops is calculated through
the exchange of bulk supergravity fields.}
\end{figure}

The coupling of the string worldsheet to the supergravity fields is
obtained by expanding the Nambu-Goto action
\eq
\label{e.ng}
\sng=\f{1}{2\pi\alp}\int dtdx e^{\Phi/2}\sqrt{\det G_{\mu\nu} \der_a
X^\mu \der_b X^\nu} -\f{1}{2}\epsilon^{ab}B_{\mu\nu} \der_a X^\mu \der_b X^\nu
\eqx
to first order in the perturbations of the background AdS fields
(denoted here generically by $\psi$):
\eq
\label{e.defvar}
\f{1}{2\pi\alp}\int \f{dtdz}{z_x} \var{\psi}(t,z) \psi
\eqx
In this work we will explicitly
consider the lightest relevant supergravity fields namely tachyonic
scalar fields (see \cite{vanNieu}) $s^I$, the 
dilaton $\Phi$, the antisymmetric tensor $B_{\mu\nu}$ and the graviton
$g_{\mu\nu}$.
Then the result is given by the integral
\eq
\label{e.form}
\f{\cor{WW}}{\cor{W}\cor{W}}=\exp\left( \f{1}{4\pi^2\alp^2} \int 4 dt_1
dt_2 \f{dz dw}{\zx \wx} \var{\psi}(t_1,z) G(x,x')
\var{\psi}(t_2,w) \right)  
\eqx
where $G(x,x')$ is the relevant (bulk-bulk) Green's function, and $x$
and $x'$ are points on the two worldsheets parametrized by $t_1,z$ and
$t_2,w$ (see figure 2).
The factor 4 in the integrand takes into account the fact that for
each value of $z$ 
there are {\em two} distinct points on the worldsheet, and that we are
performing calculations in the leading order in $a/L$ (here there is a
subtlety related to the $B_{\mu\nu}$ field which will be 
discussed in detail later on).

We note that the exponent in (\ref{e.form}) can be also interpreted as
an integral over one string worldsheet of the field $\psi(x')$
produced by the other worldsheet\footnote{It is interesting to compare
formula (\ref{e.field}) with the eikonal approximation in QCD
\cite{Kovch}, involving a one dimensional integral of the gauge field,
created by one $q\bar{q}$ state, on the Wilson loop spanned by the other.}
(with the suitable coupling $\var{\psi}(t_2,w)$) namely 
\eq
\label{e.field}
\exp\left( \f{1}{2\pi \alp}  \int 2 dt_2 \f{dw}{w_x} \var{\psi}(t_2,w)
\psi(x') \right) 
\eqx

A convenient change of variables is $v_+=t_2\sin\th$ and
$v_-=t_1-t_2\cos\th$, which gives 
\eq
\label{e.intmeas}
\int dt_1 dt_2 \lra \int \f{dv_+ dv_-}{\sin \th}\lra \int\f{rdr
d\bt}{\sin \th}
\eqx
In the last equation we chose radial coordinates for $v$'s.
The exponent in (\ref{e.form}) corresponding to the exchange of a generic
supergravity $\psi$ field will then be given by
\eq
\label{e.final}
\f{1}{4\pi^2\alp^2} \f{4}{\sin \th}\left\{ \int rdr d\bt  \f{dz dw}{\zx\wx}
\var{\psi}(z) G(x,x')\var{\psi}(w) \right\}_\psi
\eqx
where we also used the fact that by time invariance of the individual
minimal worldsheets the coupling does not depend on $t_{1,2}$. 

The $\theta$ dependence (and thus the energy dependence) is encoded
completely in the coupling and the overall jacobian $1/\sin\th$. 
The tensor structure of the
Green's function and its specific dependence on the AdS invariant $u$
(see below) gives rise to the
$a/L$ dependence of the scattering amplitude. It does not influence
the $\th$ dependence. This can be seen from the following
considerations.

The Green's function $G(x,x')$ is constructed from the invariant
bitensors\footnote{A number of useful properties are listed in
\cite{freedman}. For explicit formulas see below.} \cite{bitensor} and
scalar functions of the AdS invariant 
\eq
\label{e.u}
u=\f{(z-w)^2+\sum_{i=1}^4(x_i-x'_i)^2}{2zw}
\eqx
where $x$ and $x'$ are the two arguments of the Green's
function. In our case the 4d distance  is given by
\eq
\sum_{i=1}^4(x_i-x'_i)^2=L^2+ (t_1-t_2\cos\th)^2 +t_2^2 \sin^2\th 
\eqx
After performing the change of variables (\ref{e.intmeas}), we get
\eq
u=\f{L^2+r^2}{2 z w}
\eqx
with no $\th$ dependence.

This argument can be extended also to the tensor structure of the
Green's function. Indeed apart from functions of $u$ discussed above,
the Green's function is constructed from the bitensors
\eq
\partial_\mu u=\f{1}{z}\left[\f{(x-x')_\mu}{w}-u\dl_{\mu 0}\right]
\quad\quad
\partial_\nu' u=\f{1}{w}\left[\f{(x'-x)_{\nu'}}{z}-u\dl_{\nu' 0}\right]
\eqx
and
\eq
\partial_\mu \partial_{\nu'} u= -\f{1}{zw} \left[ \dl_{\mu\nu'} +
\f{(x-x')_\mu}{w}\dl_{\nu' 0}+ \f{(x'-x)_{\nu'}}{z}\dl_{\mu 0} -
u\dl_{\mu 0} \dl_{\nu' 0} \right]
\eqx
where $x$ and $x'$ are the two (5-dimensional) arguments of the Green's
function, $x_{\mu=0}\equiv z$ and $x'_{\nu'=0}\equiv w$.  
After the change of variables (\ref{e.intmeas}) all these quantities
do not involve $\th$. 

Therefore the energy ($\th$) dependence can be
read off directly from the coupling coefficients to the relevant
fields. One has only to verify that the appropriate Green's function 
would not give a vanishing result after integration in (\ref{e.final}).
Let us now consider the contributions of the relevant supergravity
fields.

\subsection{Scalar exchange}

Using our conventions (\ref{e.defvar}), the coupling of scalar modes
$s^I$ to the string worldsheet was derived \cite{MaldCor} to be
\eq
\var{s^I_k}=\f{-2k}{z^2}
\eqx
where the integer $k\geq2$.
The scalar Green's function has the form \cite{MaldCor}
\eq
\label{e.scprop}
\f{\al_{0,\Dl}}{B_\Dl}\cdot \f{1}{(2u)^\Dl}\;
\hyp{\Dl}{\Dl-\f{3}{2}}{2\Dl-3}{-\f{2}{u}} 
\eqx
(for the $s^I$ modes $\Dl=k$). The leading dependence in $a/L$ of
(\ref{e.final}) is governed by the asymptotic large $u$ behaviour of
the Green's function (in this case $const\cdot u^{-k}$ for large
$u$). Here and in the following we will always restrict ourselves to
calculations using this asymptotic part.
Its contribution to (\ref{e.final}) is
\eq
\biggl\{\ldots\biggr\}_{s^I_k}=k^2 \left[\f{\sqrt{\pi}c_{max}^{k-1}
\Gamma\left(\f{1+k}{4}\right)}{2\Gamma\left(\f{3+k}{4}\right)} \right]^2 \cdot
\f{\al_{0,k}}{B_k} \cdot \left(\sum_A Y_A^2\right) \cdot \f{\pi}{k-1}
\left(\f{a}{L}\right)^{2(k-1)}  
\eqx
where
\eqn
\al_{0,k}&=&\f{k-1}{2\pi^2}\\
B_k &=& \f{2^{3-k}N^2 k(k-1)}{\pi^2 (k+1)^2}\\
c_{max} &=& \f{z_{max}}{a}= \f{\Gamma(1/4)^2}{(2\pi)^{3/2}}\\
\sum_A Y_A^2 &=& \f{(k+3)(k+2)}{\pi^2 (k+1)^2}
\eqnx
where $\sum_A Y_A^2$ denotes a summation over the spherical harmonics
on $S^5$. 
After continuation to Minkowski space the whole dependence on $s$ is
\eq
\f{1}{\sin\th}=i \f{1}{\sinh \chi} \sim i\f{2m^2}{s}
\eqx
Hence the phase shift $\dl$ (see (\ref{e.dldef})) for the leading
($k=2$) mode is given by 
\eq
\dlkk=(\f{g_s}{2N} \f{10}{\pi^2})\cdot \fal^2 \cdot \f{2m^2}{s}
\eqx
The $1/s$ behaviour of $\dlkk$ is consistent with (flat space) field
theory expectations for scalar exchange. The $L$ dependence, on the
other hand, does not follow from known symmetry arguments.

\subsubsection{Dilaton exchange}

For the dilaton
\eq
\var{\Phi}=\f{1}{2}\f{\zmx^2}{z^4}
\eqx
The Green's function is analogous to (\ref{e.scprop}) but with
$\Dl=4+k$, $k\geq 0$ and the normalization $B_\Dl$ substituted by
$B'_k$ (see below). The leading contribution to (\ref{e.final}) is
\eq
\biggl\{\ldots\biggr\}_{\Phi}=\left[\f{\sqrt{\pi}c_{max}^{\Dl-1}
\Gamma\left(\f{\Dl-1}{4}\right)}{8\Gamma \left(\f{1+\Dl}{4}\right)}
\right]^2 \cdot 
\f{\al_{0,\Dl}}{B'_k} \cdot \sum_A Y_A^2 \cdot \f{\pi}{\Dl-1}
\left(\f{a}{L}\right)^{2(\Dl-1)}  
\eqx
where 
\eq
B'_k=\f{N^2}{2^{k-1} \pi^2 (k+1)(k+2)}
\eqx
This leads to the phase shift (for $k=0$)
\eq
\dl_\Phi=\f{g_s}{2N} \left(\f{192 \Gamma(5/4)^8}{\pi^8}\right)\cdot
\fal^4 \cdot \f{2m^2}{s} 
\eqx

\subsection{Antisymmetric tensor exchange}

The coupling of the antisymmetric $B$ tensor field follows from
(\ref{e.ng}).
It is therefore given by
\eq
\label{e.cfb}
\var{B_{\mu\nu}}\dl B_{\mu\nu} = \cos\th (B_{xt}+z_x B_{zt})+\sin\th(
B_{xy}+z_x B_{zy})
\eqx
However the calculation of the phase shift is in this case more
involved \cite{Mathur}. In the $AdS_5\times S^5$ background the
perturbations of the NS-NS $B_{\mu\nu}$ field mix with fluctuations of
the Ramond-Ramond 2-form
$A_{\mu\nu}$. Therefore we have to project the coupling
(\ref{e.cfb}) on to a coupling with the (lowest) mass eigenstate
$A^{eig}$, which is (a real part of a) linear combination of a 2-form
$B$ and $*dB$, where $*$ is the Hodge dual. Therefore this state
should couple \cite{Mathur} to the string through a linear combination
of the coupling (\ref{e.cfb}) and the derivative coupling:
\eq
\label{e.cfbdual}
\var{B_{\mu\nu}} \cdot \sqrt{G} \cdot
\mbox{$\epsilon_{\mu\nu}$}^{\rho\sg\dl}\partial_\rho A^{eig}_{\sg\dl} 
\eqx
We will not determine here the appropriate coefficients and
normalization but rather concentrate on isolating the $a/L$ and $s$
dependence which is relevant for our analysis. 

We will begin by considering the nonderivative coupling (\ref{e.cfb})
with the substitution $B_{\mu\nu}\rightarrow A^{eig}_{\mu\nu}$

The Green's function $G_{\mu\nu\mu'\nu'}$ should be constructed from
the invariant bitensors and should be antisymmetric in both pairs of
indices. These requirements lead to the following tensor structures:
\eqn
T_1^B\!\! &=& \!\!\uzw{\mu}{\mu'}\uzw{\nu}{\nu'}-\uzw{\nu}{\mu'}\uzw{\mu}{\nu'}\\
T_2^B\!\! &=& \!\! \uzw{\mu}{\mu'}\du{\nu} \du{\nu'}-\uzw{\nu}{\mu'}\du{\mu} \du{\nu'}
-\uzw{\mu}{\nu'}\du{\nu} \du{\mu'}+\nonumber\\
&&+\uzw{\nu}{\nu'} \du{\mu} \du{\mu'}
\eqnx
The Green's function is then given\footnote{Here we consider the
lowest mode $k=1$ of the $m^2=ek^2$ family of \cite{vanNieu}.} for
large $u$ \cite{Mathur} to be
\eq
G^B=(Normalization)\left(-\f{1}{u^3}T_1^B +\f{1}{u^4}T_2^B\right)
\eqx

Here we encounter the subtlety mentioned after formula (\ref{e.form}).
The contraction of the Green's function with the coupling (\ref{e.cfb})
has to be done with care, as there are a number of terms which are
linear in the derivatives $z_x$. When we integrate the resulting
expression over the string worldsheet, and we change variables to $z$,
we have to recall that there are {\em two} values of $x$ corresponding
to each $z$ so we have in effect:
\eq
\int_0^a dx I\lra \int_0^{\zmx}\f{dz}{\zx} \left(I+I_{\zx \ra -\zx, x \ra
a-x}\right)
\eqx
Therefore the contribution would vanish unless we also keep the
full expression (\ref{e.u}) for $u$ i.e.
\eq
u=\f{(L+x_1-x_2)^2+(z-w)^2+(t_1-t_2)^2}{2zw}
\eqx
After some computer algebra the result (leading in $a/L$) is
\eq
\cor{\var{B_{\mu\nu}}\dl B_{\mu\nu} \var{B_{\mu\nu}}\dl B_{\mu\nu}} =
(number)\f{32\pi w z}{3L^4}\cos\th 
\eqx
Performing the remaining integration leads to
\eq
\dlb=\f{g_s}{2N}\cdot (number')\cdot \left(\f{a}{L}\right)^4  
\eqx
The leading energy dependence follows from the angular behaviour
\eq
\f{\cos\th}{\sin\th}=i\coth\chi\sim is^0
\eqx
This behaviour $i \coth\chi$ is exactly the same which appears in
field theory with vector exchange (\ref{e.qedeik}).

Let us now go back and consider the derivative coupling
(\ref{e.cfbdual}). By the arguments in section \ref{s.ww} the leading 
$\th$ dependence will not change. Moreover an explicit calculation
shows that the leading contribution to the phase shift has exactly the
same $(a/L)^4 \coth \chi$ dependence as for the coupling
(\ref{e.cfb}). In the high energy limit we have therefore a
contribution of the 2-form field:
\eq
\dlb=\f{g_s}{2N}\cdot (number'')\cdot \left(\f{a}{L}\right)^4 
\eqx

\subsection{Graviton exchange}

The graviton coupling can be obtained by expanding the Nambu-Goto
action and retaining the first order:
\eq
\label{e.gravcp}
\sqrt{h}\left[ \f{\dl\gtt}{2\gtt} +\f{\dl\gss}{2\gss} \right] =
\f{\zmx^2}{2z^2}\dl\gtt+\f{z^2}{2\zmx^2}\dl\gss
\eqx
and the expression for the variation of the induced metric
$h_{ab}=G_{\mu\nu} \partial_a X^\mu \partial_b X^\nu$ in terms of
the bulk metric:
\eqn
\label{e.htt}
\dl \gtt^\th &=&  \cos^2\th \dl G_{tt} +\sin^2\th \dl G_{yy}+ \sin2\th \dl
G_{ty}\\
\label{e.hss}
\dl \gss &=& \dl G_{xx} +z_x^2 \dl G_{zz}+ 2z_x \dl
G_{zx}
\eqnx
The graviton field $g_{\mu\nu}\equiv \dl G_{\mu\nu}$ should be
decomposed into a scalar 
part $g^\al_\al$ that mixes with the RR field strength and a `pure'
graviton field $g'_{\mu\nu}$:
\eq
\label{e.hdecomp}
g_{\mu\nu}=g'_{\mu\nu}-\f{1}{3}G_{\mu\nu} g^\al_\al
\eqx
 
The physical graviton propagator (for
$g'_{\mu\nu}$) is (see \cite{Freedman}) 
\eq
\label{e.gravgr}
\kappa^2 \left[(\der_\mu\der_{\mu'}u \der_\nu\der_{\nu'}u+
\der_\mu\der_{\nu'}u \der_\nu\der_{\mu'}u) G(u)+g_{\mu\nu}
g_{\mu'\nu'} H(u)\right]
\eqx
with $G(u)\sim (3/32\pi^2)\cdot 1/u^4$, $H(u)\sim (-1/48\pi^2) \cdot
1/u^2$ for large $u$. $\kappa^2$ is the gravitational constant equal
in our units $\kappa^2=15\pi^3/(2N^2)$.
Explicit expressions for $G(u)$ and $H(u)$ are given in \cite{Freedman}. 

Using this expression one can explicitly check that the whole $\th$
dependence arises from the correlation function $\cor{\dl\gtt^\th
\dl\gtt^{\th=0}}$, defined by contracting (\ref{e.htt}) with the
Green's function (\ref{e.gravgr}). 
Residual $\th$ dependence in $\cor{\dl\gtt^\th \dl\gss}$
cancels after performing the angular $\bt$ integral in
(\ref{e.final}). Explicitely we get
\eqn
\!\!\!\!\!\!\!\!&&\cor{\dl\gtt^\th \dl\gtt^{\th=0}} =\f{1}{z^2
w^2}\left[2G(u)\cos^2\th
+H(u) \right]\nonumber\\
\!\!\!\!\!\!\!\!&&\cor{\dl\gtt^\th \dl\gss} = \f{1}{z^2 w^2}\left[\f{2 r^2
\zx^2 \cos^2(\bt+\th)}{z^2}G(u) +(1+\zx^2)H(u) \right]\nonumber\\
\!\!\!\!\!\!\!\!&&\cor{\dl\gss \dl\gss}=\f{1}{z^2 w^2}\biggl[
\f{2(w^2\wx\zx+\wx z(x_1-x_2+z\zx)+}{
w^2z^2}\nonumber\\
&&\quad\quad\quad\quad\f{+w(z-(x_1-x_2+(1+u)\wx z)\zx))^2}{
w^2z^2} \cdot G(u)+\nonumber\\
&&\quad\quad\quad\quad +(1+\wx^2)(1+\zx^2) H(u)\biggr]
\eqnx
Inserting this into (\ref{e.final}) and using (\ref{e.gravcp}) we get
the result 
\eq
\label{e.gravamp}
\f{1}{4\pi^2\al'^2} \kappa^2 \f{c_{max}^6\Gamma^2\left(\f{3}{4}\right)}{
\Gamma^2\left(\f{1}{4}\right)}
\left(\f{a}{L}\right)^6
\f{\cos^2\th}{\sin\th}+\f{1}{\sin\th}\cdot(\mbox{$\th$-independent pieces}) 
\eqx
The leading energy behaviour now follows from:
\eq
\f{\cos^2\th}{\sin\th}=i\coth\chi \cosh\chi\sim i\f{s}{2m^2}
\eqx
So the phase shift is 
\eq
\dlgrav=\f{g_s}{2N} \f{15\pi^3}{2} \f{c_{max}^6\Gamma^2\left(\f{3}{4}\right)}{
\Gamma^2\left(\f{1}{4}\right)}
\left(\f{a}{L}\right)^6 \cdot \f{s}{2m^2}
\eqx
This is a rather unexpected prediction for a field theory scattering
amplitude. Before we discuss in more detail the range of validity of
the above expression we will first analyze whether the counterterms introduced
by Drukker, Gross and Ooguri \cite{neww}, which are necessary for the
finiteness of the Wilson loop expectation values, can change the above
result.

\section{The Legendre transform prescription}

In \cite{neww} the Wilson loop prescription was modified by taking the
Legendre transform:
\eq
\label{e.modif}
\atl\lra A-\int d\tau \sqrt{h} h^{\sg\sg}G_{ij}Y^i \dx Y^j
\eqx
where $A$ is the Nambu-Goto action while $Y^i=z\Theta^i$ and the
$\Theta^i$ are coordinates on $S^5$ satisfying $\sum_{i=1}^6
\Theta^i\Theta^i=1$. The boundary contribution in (\ref{e.modif})
cancels exactly the $1/z$ divergence arising from integrating (\ref{e.area}). 
Since the additional term does depend on the
metric, it might, in principle, change (\ref{e.gravamp}). It suffices
to calculate it's behaviour under variations of $\gtt$. We get
\eq
\label{e.nocontr}
\int d\tau \sqrt{h}\f{1}{2\gtt} h^{\sg\sg}G_{ij} Y^i \dx Y^j \dl\gtt=
\int d\tau \f{z^4}{2\zmx^2} \f{z_x}{z}\dl\gtt\sim \int d\tau z \dl\gtt
\eqx
which vanishes when $z\lra 0$ (note that the graviton Green's
functions are nonsingular at the boundary --- they vanish).

However one could envisage a modification of the bulk action which
would regularize the action upon imposing
equations of motion. Suppose, for instance, that we modify the action to
\eq
\sqrt{h}(1-c_1)+\dx(\sqrt{h})c_2
\eqx
where the counterterms $c_1$ and $c_2$ do not depend explicitely on $\gtt$
(e.g. $g^{ab} G_{ij} \partial_a Y^i \partial_b Y^j$ is such a term).
This is equivalent to
\eq
\label{e.arg}
\sqrt{h}(1+c_1-\dx c_2)+\dx(\sqrt{h} c_2)
\eqx
For the second term to cancel the Nambu-Goto singularity arising from
$\sqrt{h}\sim z^{-4}$, $c_2$ should behave
like $z^3$ and hence by similar reasoning as in (\ref{e.nocontr}) it
would lead to a vanishing coupling to the graviton and so would
not give a contribution. Any `softer' behaviour like $z^2$ would, on
the other hand, introduce additional singularities which would have to
be cancelled by additional counterterms etc.
Therefore the cancelation of the $\dl\gtt$
coupling should come just from the first term in (\ref{e.arg}). But
now all the coupling to $\dl\gtt$ comes from $\sqrt{h}$ so the term in
parentheses should give 0 upon inserting equations of motion. 

But clearly this is impossible, since this would lead to wrong results
for $\cor{W}$. The above 
argument does not prove, of course, that some more complicated
counterterms would not help to cure this behaviour, but it does not seem
to be very likely.

\section{Range of validity of the weak field approximation}

In this section we will analyze the range of validity of the results
obtained above. 
The general assumption in these calculations is that the impact
parameter $L$ is much greater than the transverse size $a$ of the
$W$-boson pair (which plays the role of a quark-antiquark pair).
A second constraint\footnote{We are grateful to J. Maldacena for
pointing this out.} comes from the fact that we are staying within the
linearized gravitational regime.  
We will
concentrate on the graviton contribution which leads to the strongest
constraint. The calculations should be valid when the field $\dl
G_{tt}$ produced by one of the moving (tilted) worldsheets, evaluated
at points on second worldsheet,  should be
much smaller than the background AdS metric $G_{tt}$.
\eq
\label{e.gttc}
\dl G_{tt}\ll G_{tt}=\f{1}{w^2}
\eqx
The field produced at the point $t_1=0$ by the second (tilted)
worldsheet is given by (compare (\ref{e.form}) and (\ref{e.field}))
\eq
\f{1}{2\pi\al'}\int dx dt \f{\zmx^2}{2z^2} \cor{\dl\gtt^\th
\dl\gtt^{\th=0}}
\eqx
Keeping only the $\th$-dependent term which is dominant after making
the transition to Minkowski space, and using $u=(L^2+t^2)/(2zw)$
one gets
\eq
\f{1}{2\pi\al'}\,2\int\f{dz}{z_x} \f{\zmx^2}{2z^2}\f{1}{z^2 w^2}
2\cos^2\th\, \f{3}{32\pi^2} \int_{-\infty}^\infty \f{2^4 z^4
w^4}{(L^2+t^2)^4} dt
\eqx
Which gives finally
\eq
(number)\f{w^2 a^3}{L^7}\cos^2\th \ll \f{1}{w^2}
\eqx
The constraint is most restrictive when evaluating at $w\sim a$, which
is as far as the string worldsheet extends into the
$5^{th}$ dimension of the $AdS_5$, so we
get finally (dropping factors of order 1, and taking some unit mass
$m\sim 1$):
\eq
\label{e.constraint}
\cos^2\th\sim s^2 \ll \left(\f{L}{a}\right)^7
\eqx
We see that when we fix the impact parameter we cannot go to arbitrarily
high energies. At some point the gravitational field becomes so strong
that one would have to consider multigraviton exchanges and presumably
perform resummation using the full picture of strings propagating in
AdS space. The analysis of this regime is beyond the scope
of this paper. We note that this constraint is specific to the
analytic continuation from Euclidean to Minkowski geometry.

It is to be noticed that when (\ref{e.constraint}) is satisfied the
graviton phase shift $\dlgrav$ is indeed small as it should be
\eq
\label{e.grph}
\dlgrav\sim \left(\f{a}{L}\right)^6 s \ll \left(\f{a}{L}\right)^{6-7/2=2.5}
\eqx
Other fields considered by us lead to less stringent constraints.

\section{Analysis of the high energy scattering amplitudes at large
impact parameter}

The leading $L$ and $s$ dependence of the phase shifts for the
relevant supergravity fields is $\dlkk=\propto (a/L)^2 s^{-1}$ for the
tachyonic $s^I$ scalar, $\dlb=\propto(a/L)^4$ for the $B_{\mu\nu}$ field
and $\dlgrav=\propto (a/L)^6 s$ for the graviton. The proportionality
constants, obtained explicitly for the scalars and the graviton, are real.

The fact that we obtained real amplitudes is linked with the
expectation that for large impact parameters the scattering should be
predominantly elastic. 

Standard crossing relations and analyticity properties of amplitudes
relate the phase of the amplitude with the energy behaviour and
signature of the exchanged state. For amplitudes behaving like $1/s$
and $s$, a real amplitude implies positive signature $\xi=+1$, while for the
constant amplitude of the antisymmetric tensor we should have negative
signature $\xi=-1$. This is indeed consistent with the behaviour of the 
coupling of these fields to the string worldsheet under an exchange of
the quark with the antiquark in one of the Wilson loops. This exchange
translates into a change of orientation of the associated
string worldsheet. The coupling to the scalars and the graviton stays
invariant ($\xi=+1$) while the coupling to the $B_{\mu\nu}$ field
changes sign. In this way we may also separate off the $B_{\mu\nu}$
contribution from the others. 

Now we would like to analyze which supergravity fields
are dominant in different regions of the $(a/L,s)$ parameter plane.
We stay within the weak field approximation (\ref{e.constraint}) where
all the phase shifts are small.
We will consider two regimes.

\subsubsection*{Fixed $n\equiv\log\f{L}{a}/\log s$}

Because of the form of the constraint (\ref{e.constraint}) it will turn
out to be convenient to parameterize
\eq
\f{L}{a}=s^n
\eqx
where $n$ is a (real parameter), and look for dominant contributions
when increasing $s$, while keeping $n$ fixed. The constraint now is
just $n>2/7$. The $n$ dependence of the phase shifts for the tachyonic
scalar, antisymmetric tensor field and the graviton
is in this
parameterization $\dlkk\sim s^{-1-2n}$, $\dlb\sim s^{-4n}$ and
$\dlgrav\sim s^{1-6n}$ respectively. There are 3 cases.
For $n$ between $2/7$ and $1/2$, the graviton
contribution dominates, followed by $B_{\mu\nu}$ and the KK $s^I$
scalar, at $n=1/2$ all 3 contributions are comparable, 
the precise values of the numerical coefficients (ignored in this
analysis) would eventually differentiate between the 3 contributions
which all behave here like $(a/L)^4$, while for
$n>1/2$ the KK $s^I$ scalar dominates. 
Notably enough, the above shows that in a certain region of (large
impact parameter) phase space the bulk
graviton exchange gives a dominant contribution to the SYM scattering
amplitude. 

\subsubsection*{Fixed $L$}

When we keep $L/a$ fixed and large and increase $s$ (but staying below
the limit $(L/a)^{7/2}$) we find that the graviton contribution gives
a linear rise of the phase shift with $s$. Indeed although $s$ is not
arbitrarily large it dominates over the $B_{\mu\nu}$ contribution:
\eq
\dlgrav\sim \left(\f{a}{L}\right)^6 s\sim  \left(\f{a}{L}\right)^2 s
\cdot \dlb \sim \left(\f{L}{a}\right)^{7/2-2=1.5} \cdot \dlb
\eqx
So for sufficiently large $L/a$ we see a linear rise of the phase
shift with $s$ in gauge theory.
This behaviour is rather unexpected in SYM (see the discussion).
Note however, that within the constraints, we are in the region of
applicability of the eikonal approximation as the graviton phase shift
(\ref{e.grph}) remains small.

\section{Discussion}

At this point we may qualitatively contrast the situation with the
case of perturbative (non-supersymmetric) QCD. The leading large log
resummation give rise \cite{BFKL} to amplitudes rising like
\eq
s^{\f{4}{\pi}\al_s N \log 2}
\eqx
The exponent depends on the coupling constant, and increases with
$\al_s$ --- this is a dynamical effect coming from enhanced gluon
radiation at small $x$. In the strongly coupled phase (our
calculation) we get the exponent $1$ which is purerly `kinematical'
--- it reflects just the spin-2 nature of the graviton. This may
perhaps be thought of as a nonperturbative strong coupling limit of some
dynamically generated enhancement, but if it were so there still remain some
questions.

On the perturbative side one expects unitarity effects to set in at
large $s$ thus leading to a constant $s^0$ or at most logarithmic
$\log s$ behaviour of the phase shift\footnote{However there may be
subtleties in explicitly carrying over of the Froissart bound to the
${\cal N}=4$ SYM, nonconfining CFT case.}. We may postulate that the
same kind of unitarization {\em on the supergravity side}, might bring down the
behaviour from $s^1$ to $s^0$. However this would involve some very subtle
behaviour specific to AdS. Qualitatively such a strong suppression
would call for some cancelation of gravity in the high energy
limit. The RR force that cancels the static gravitational interaction between
D-branes \cite{Polchinski} is negligible when one goes to the high
relative velocity limit \cite{Bachas}. One can check that indeed in
the large impact parameter limit the dominant contribution to the
D-brane scattering phase shift comes also from single graviton
exchange leading to the phase shift proportional to $s$. In the same
regime the contribution of the excited string modes would be
supressed. We expect the same pattern of behaviour here.

For the case of 4-point Virasoro-Shapiro amplitude, which contains the
contribution of higher string modes, the large $s$, fixed $t$ regime
gives just a modification of the gravity tree level result due to reggeization:
\eq
\f{s^2}{t}\lra \f{s^2}{t}\cdot \exp\left[\f{t\al'}{2}\log(\al' s)\right]
\eqx
This is a slightly wider regime than the one considered by us, as
fixing even small $t$ involves integration over all impact parameters.
The phase shift for small $t$ is again proportional to $s$.

Also analysis of perturbative resummation in gravity and string theory
in flat space (see e.g. \cite{Amati}) does not seem to soften the
energy behaviour.

For the case of $AdS_5$ one expects some differences, in particular 
one obtains a different $t$ dependence without the $1/t$ pole (which is
regulated by the finite radius\footnote{In the large radius
limit the dominant contribution to the Green's functions should come
from the region of small $u$. All the above Green's functions behave
like $u^{-3/2}$, which translates into $1/R^3$ behaviour
characteristic of a free scalar propagator in flat 5D space.} of
AdS). It would be extremely interesting to investigate these issues in the AdS
context, however as we saw, the tree level energy behaviour had exactly
the same origin as in flat space.

The $s$ behaviour obtained by us seems to be quite
a robust and generic feature of gravity mediated scattering in various
contexts. However the novel feature of the AdS/CFT correspondence
allows to translate this kind of behaviour to gauge theory scattering
amplitudes. This makes the full understanding of
the high energy behaviour for AdS string theory into an
interesting and subtle problem.

\section*{Acknowledgements} 
One of us (RJ) would like to thank Juan
Maldacena and Maciej A. Nowak for discussions. We are grateful to 
Costas Bachas for a helpful discussion.
We thank Samir Mathur for
communicating to us his results on the $B_{\mu\nu}$ propagator prior
to publication. RJ was supported in part by KBN grants 2P03B01917 and
2P03B00814.

\medskip

{\em Note added:} As the final version of this paper was being typed a
preprint \cite{Zahed} appeared which addresses a related problem of
high energy scattering of (coloured) quarks.

\end{document}